\documentclass{IEEEcsmag}
\usepackage{xcolor}
\usepackage[normalem]{ulem}

\usepackage[colorlinks,urlcolor=blue,linkcolor=blue,citecolor=blue]{hyperref}
\expandafter\def\expandafter\UrlBreaks\expandafter{\UrlBreaks\do\/\do\*\do\-\do\~\do\'\do\"\do\-}
\usepackage{upmath, color}
\usepackage{booktabs}
\usepackage{siunitx}
\usepackage{orcidlink}

\jvol{XX}
\jnum{XX}
\paper{8}
\jmonth{\tiny{Preprint. This article has been accepted for publication in IEEE Reliabiltiy Magazine. This is the author's version which has not been fully edited and content may change prior to final publication.}}
\jname{}
\jtitle{}
\pubyear{}

\begin{document}

\sptitle{Article Type: Original Article}

\title{Taming Silent Failures: A Framework for Verifiable AI Reliability}

\author{Guan-Yan Yang$^{\orcidlink{0009-0002-2539-9057}}$}
\affil{National Taiwan University, Taipei, 106319, Taiwan R.O.C}

\author{Farn Wang$^{\orcidlink{0000-0002-0425-6500}}$}
\affil{National Taiwan University, Taipei, 106319, Taiwan R.O.C}


\markboth{FEATURE}{FEATURE}

\begin{abstract} \sloppy{\looseness-1
The integration of Artificial Intelligence (AI) into safety-critical systems introduces a new reliability paradigm: silent failures, where AI produces confident but incorrect outputs that can be dangerous. This paper introduces the Formal Assurance and Monitoring Environment (FAME), a novel framework that confronts this challenge. FAME synergizes the mathematical rigor of offline formal synthesis with the vigilance of online runtime monitoring to create a verifiable safety net around opaque AI components. We demonstrate its efficacy in an autonomous vehicle perception system, where FAME successfully detected 93.5\% of critical safety violations that were otherwise silent. By contextualizing our framework within the ISO 26262 and ISO/PAS 8800 standards, we provide reliability engineers with a practical, certifiable pathway for deploying trustworthy AI. FAME represents a crucial shift from accepting probabilistic performance to enforcing provable safety in next-generation systems.}
\end{abstract}

\maketitle

\chapteri{A}{rtificial} Intelligence (AI) has become a core building block of autonomous and intelligent systems. From driver assistance to computer-aided diagnosis (CAD), data-driven components promise superhuman perception and decision support. Yet they also introduce a reliability problem that differs from classical, code-centric software engineering: \textbf{silent failure}, confident outputs that are wrong, with no explicit crash, exception, or error code exposed to the rest of the stack \cite{hendrycks2018baselinedetectingmisclassifiedoutofdistribution,silentfailure2023}.

Safety-critical traditional software is developed under rigorous processes (requirements traceability, design assurance, redundancy, and diagnostics) and can exhibit multiple failure modes (e.g., fail-silent, latent, Byzantine), which are analyzed and mitigated through established standards and verification activities. In contrast, the correctness of learning-enabled components depends on data distributions as much as on code, and can degrade under distribution shift, sensor faults, or occlusions without tripping conventional diagnostics \cite{hendrycks2018baselinedetectingmisclassifiedoutofdistribution}. This motivates a runtime, contract-based approach that keeps system behavior within a provably safe envelope even when Machine Learning (ML) components behave unexpectedly.

Standard testing is insufficient, as the input space of production DNNs is hyper-dimensional and cannot be exhaustively exercised \cite{zhu2023verification}. We therefore advocate continuous, formally grounded runtime assurance of the \emph{system} rather than attempting to fully verify the internals of the AI model.

This article argues for a necessary paradigm shift in how we approach the reliability of AI-enabled systems. We must transition from a futile attempt to achieve perfect pre-deployment validation of the AI model itself to a strategy of continuous, formally grounded runtime assurance of the entire system's behavior. We introduce the \textbf{F}ormal \textbf{A}ssurance and \textbf{M}onitoring \textbf{E}nvironment (FAME), a framework designed to wrap a verifiable safety net around complex, untrustworthy AI components. FAME leverages the mathematical precision of formal methods to automatically synthesize lightweight, high-speed runtime monitors from system-level safety requirements. These monitors act as independent, real-time arbiters of the AI's behavior, checking its observable inputs and outputs against inviolable safety rules. By moving assurance from a purely offline activity to a hybrid offline-synthesis and online-enforcement model, FAME provides a practical, powerful, and certifiable methodology for building verifiably reliable AI systems. This paper will detail the FAME architecture, demonstrate its efficacy through a rigorous proof-of-concept, and situate it within the broader context of industrial safety standards and future research directions.

\section{Positioning against the State-of-the-Art}
Current research in AI safety predominantly follows two paths. FAME carves out a third, more pragmatic route.

\subsection{White-Box Formal Verification of DNNs}
Efforts to prove properties of network internals and outputs (such as local robustness) provide strong, pre-deployment guarantees but struggle with scale and property expressiveness for production models \cite{cheng2023runtime}. These methods remain valuable where tractable, but cannot replace runtime defenses in open-world operation.

\subsection{Robustness, Out-of-Distribution (OOD), and Uncertainty.}
Adversarial training and data augmentation improve statistical robustness \cite{yang2025resilience}; OOD detectors and uncertainty estimators can suppress or gate low-confidence outputs \cite{hendrycks2018baselinedetectingmisclassifiedoutofdistribution}. These are model- and dataset-dependent and do not give deterministic, requirement-centric guarantees. FAME complements them by enforcing time-bounded contracts on observable signals, independent of internal calibration or training regimen.

\subsection{Runtime Verification and Assurance}
Runtime verification monitors executions against formal specifications in operation \cite{bartocci2018survey}. Run-time assurance architectures often switch between \textbf{advanced and safe} controllers \cite{cofer2020run}. FAME specializes these ideas to perception- and data-driven components by: (i) a specification engineering workflow for continuous-valued signals using bounded-future STL; (ii) a code generator producing deterministic, constant-overhead online monitors; and (iii) a feedback loop that curates violation contexts for model retraining and property refinement.

\section{The FAME Framework: A Two-Phase Architecture for Continuous Assurance}

\begin{figure}[!t]
\centering
\includegraphics[width=\columnwidth]{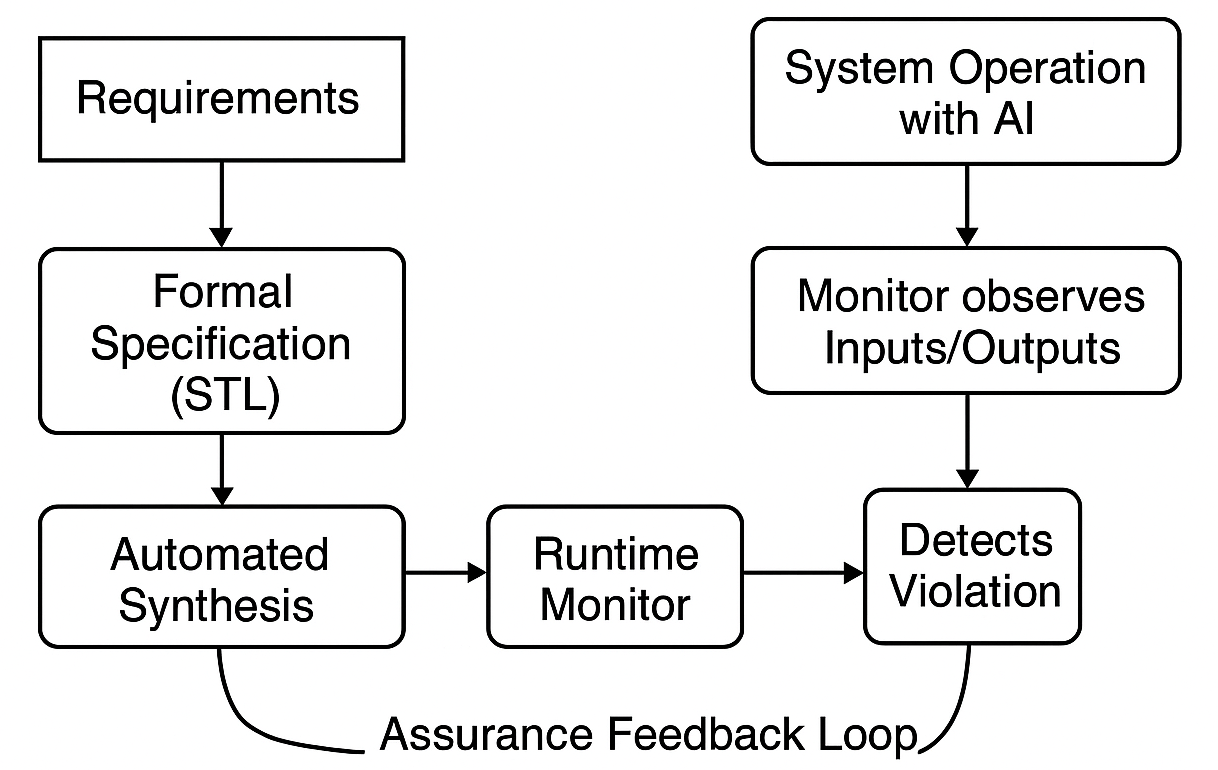}
\caption{The FAME framework integrates offline synthesis of monitors from formal specifications with online enforcement and a feedback loop for continuous assurance.}
\label{fig_framework}
\end{figure}

The FAME framework establishes a continuous assurance lifecycle that bridges the gap between system design and operation. It is founded on a key philosophical principle: instead of attempting the often-intractable problem of formally verifying the internal state of a massive DNN, we should focus on verifying its observable behavior against a formally specified contract. This black-box approach is key to its scalability and model-agnostic nature. The framework consists of two primary phases: Design-Time Synthesis and Run-Time Mitigation, linked by a crucial feedback loop that enables system evolution and learning, as illustrated in Fig. \ref{fig_framework}.

\subsection{Phase 1: Design-Time Specification \& Synthesis}
The foundation of FAME is a verifiable and correct-by-construction \cite{sun2025correct} safety mechanism.

\subsubsection{Formal Specification: The Language of Safety} 
The process begins by capturing critical safety and functional requirements in a precise, unambiguous mathematical language. Natural language is notoriously ambiguous; a requirement like "The car should detect pedestrians in front of it" is useless for verification. We must be precise about distance, timing, and confidence. For systems with dynamic, real-time behavior, \textbf{Signal Temporal Logic (STL)} is effective \cite{wang2004formal}: it provides dense-time temporal operators over real-valued signals and admits quantitative (robust) semantics. Alternatives include linear temporal logic/metric temporal logic for discrete events and metric timing, timed logics with explicit time variables, and stream-processing logics/languages (such as those surveyed in \cite{bartocci2018survey}) that are well suited to event streams. We choose STL because (i) the signals of interest (distances, velocities, confidences) are continuous-time/real-valued; (ii) robust semantics enable margin-aware triggers; and (iii) efficient online monitors can be synthesized under bounded horizons.

These formal specifications form the bedrock of the framework and are typically authored by systems engineers and safety experts, rather than ML scientists. They can be derived from several sources:
\begin{itemize}
    \item \textbf{System Requirements:} Directly translating safety goals from standards like ISO 26262 \cite{iso26262_2018} and ISO/PAS 8800 \cite{iso8800_2024} .
    \item \textbf{Domain Expertise:} Capturing heuristics from experienced operators (e.g., "A robotic arm's velocity must decrease as it approaches its target").
    \item \textbf{Data-Driven Mining:} Using techniques like temporal specification mining to automatically discover invariant properties from large logs of simulation data \cite{wang2013temporal}.
\end{itemize}

Table \ref{tab:spec_comparison} illustrates the value of this precision.

\begin{table*}[!t]
\caption{Comparison of Requirement Specification Types. Note that $G$ means 'always' (Globally), $\land$ means 'AND', $\implies$ means 'IMPLIES', and $F_{[a, b]}$ means 'eventually' (Finally) within the time interval $[a, b]$.}
\label{tab:spec_comparison}
\centering
\begin{tabular}{@{}ll@{}}
\toprule
\textbf{Type} & \textbf{Example Requirement} \\ \midrule
Natural Language & Detect pedestrians reliably. \\ \addlinespace
Vague & Confidence should be high for nearby pedestrians. \\ \addlinespace
\textbf{Formal (STL)} & $G ((\text{dist} < 30 \land \text{is\_ped}) \implies F_{[0, 0.1]} (\text{conf} > 0.8))$ \\
& \small{\textit{(Always, if dist is <30m and obj is a pedestrian, then}} \\
& \small{\textit{within 0.1s, confidence must exceed 0.8.)}} \\ \bottomrule
\end{tabular}
\end{table*}
An STL formula can constrain how signals should evolve over time, making it ideal for specifying requirements like "the brake must be applied within \SI{100}{\milli\second} of a hazard detection" or "the confidence in a pedestrian classification must not drop below 0.8 while the object is within \SI{30}{\meter}." This formalism removes ambiguity and creates a testable, verifiable contract.

\subsubsection{Specification Engineering and Proactive Stressing}
To reduce specification gaps, we institute a design-time loop:
\begin{enumerate}
\item \textbf{Hazard Analysis to Properties:} Starting from safety goals (ISO 26262) and AI-specific risks (ISO/PAS 8800), we derive candidate properties with explicit time and magnitude bounds.
\item \textbf{Proactive Stressing:} We generate counterexample-seeking scenarios via (i) fault injection on sensors (dropouts, saturation, bias), (ii) environment stressing (rain, fog, glare, occlusion), and (iii) distribution-shift perturbations. Each scenario attempts to falsify current properties while remaining within the Operational Design Domain (ODD).
\item \textbf{Property Refinement:} Violations are triaged as specification shortcomings (false alarms) or true hazards (missed by AI). Thresholds and windows are adjusted to minimize false positives while preserving detection coverage.
\end{enumerate}
This loop complements data-driven training by directly testing the sufficiency of the \emph{contract}, not only the model.

\subsubsection{Controlled NL-to-STL Translation}
We use controlled templates to ensure traceability.

\noindent\textbf{Template 1 - Timely Response:}
"If $C$ holds, then within $\Delta t$, $R$ must hold":
\[
G(C \rightarrow F_{[0,\Delta t]} R).
\]

\noindent\textbf{Template 2 - Sustained Confidence:}
"While in region $Q$, confidence must be above $\gamma$ for at least proportion $\rho$ over window $\tau$":
\[
G\left(\text{in}Q \rightarrow \mathsf{Prop}{\rho,W}(\text{conf} \ge \gamma)\right),
\]
with $W=\lceil f_s\tau\rceil$ and $\mathsf{Prop}_{\rho,W}$ implemented by counting satisfactions in a ring buffer.

\subsubsection{Automated Monitor Synthesis Toolchain}
We implement code generation with RTAMT \footnote{This library is available in \href{https://github.com/nickovic/rtamt}{https://github.com/nickovic/rtamt}.}, a runtime verification library for quantitative temporal properties. STL predicates compile to inline evaluators; bounded-time operators allocate ring buffers sized by their horizons, yielding $O(1)$ work per predicate per sample and memory $O(H)$ for horizon $H$. The generator emits:
(i) a portable C/C++ library; (ii) a ROS~2/DDS node that subscribes to typed topics and publishes a Boolean violation flag plus robustness margin; (iii) auto-documentation pairing each STL property with its controlled-English paraphrase. Semantic correctness is validated by cross-checking monitor outputs against offline evaluation. This supports semantics‑preserving code generation validated against offline oracles.

\subsection{Phase 2: Run-Time Monitoring \& Mitigation}
During operation, the lightweight, synthesized monitors are deployed alongside the AI component, adding a real-time assurance layer.

\subsubsection{In-Situ Monitoring: The Watchful Eye} 
The monitors non-intrusively observe the data streams flowing into and out of the AI model. In a modern automotive or robotic stack, this involves subscribing to topics on a data bus like DDS or ROS. For a perception system, the monitor would observe the raw sensor feeds, the AI's object classification labels, confidence scores, and bounding box coordinates. The monitor's state is updated with every new data frame, continuously evaluating the STL formula against the real-world data stream. This approach is a practical application of the principles of runtime verification, which has emerged as a key technique for assuring complex systems where exhaustive offline verification is impossible \cite{cheng2023runtime}. 

\subsubsection{Violation Detection \& Mitigation Strategy Activation} 
If the AI's behavior ever contradicts the formal specification, the monitor instantly detects the violation. This detection is not a system failure; it is a successful activation of the safety net. The monitor's output is a simple binary signal: `true` (violation) or `false` (compliant). This signal is then used to trigger a pre-defined and independently verified mitigation strategy. The choice of strategy is critical and application-specific:
\begin{itemize}
    \item \textbf{Fail-Safe:} The highest level of safety. Upon violation, the system transitions to a minimal risk condition (MRC). For a vehicle, this could be a controlled emergency stop. For a medical device, it could mean halting treatment and alerting a clinician.
    \item \textbf{Fail-Operational:} For systems requiring high availability, the monitor can trigger a switch to a redundant component. This could be a second, diverse AI model or, more robustly, a simpler, non-AI-based backup controller that provides minimal but safe functionality.
    \item \textbf{Fail-Degraded:} In less critical situations, the system might degrade its performance to a safer state. For example, an autonomous vehicle might reduce its speed, increase its following distance, and alert the driver to take over.
\end{itemize}

\subsection{Macro-Explainability Payload}
\begin{figure*}[!ht]
\centering
\includegraphics[width=1.75\columnwidth]{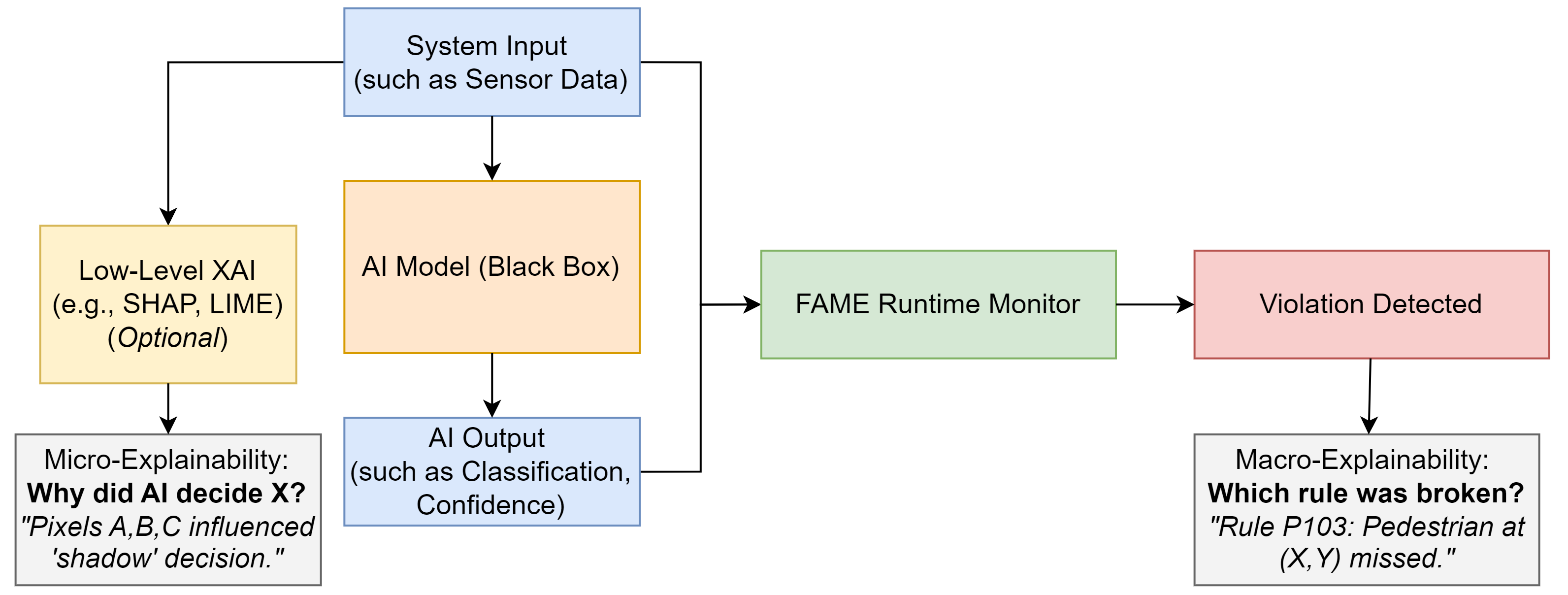}
\caption{FAME's macro-explainability versus low-level XAI. While low-level XAI seeks to explain \textit{why} an AI made a specific decision (e.g., pixel importance), FAME's macro-explainability definitively answers \textit{which system-level safety rule was violated}, providing actionable intelligence for reliability engineers.}
\label{fig_explainability}
\end{figure*}

Low-level XAI (such as SHAP, LIME) explains model internals but may be weakly actionable for operators \cite{kuznietsov2024explainable,atakishiyev2025safety}. FAME provides macro-explanations tied to system-level rules: 
rule identifier and controlled-English text, robustness margin $r(t)$, culprit signals over an evidence window, and a minimal repair suggestion. This payload is designed first for machines, such as automated Machine Learning Operations (MLOps), then optionally for humans. Fig.~\ref{fig_explainability} contrasts macro- versus micro-level explainability.
Payloads support immediate action, curated retraining, and audit trails for ISO 26262 and ISO/PAS 8800 processes.
Crucially, this machine-readable payload can serve as a direct trigger for deeper, automated root-cause analysis, invoking low-level XAI methods, such as SHAP or saliency mapping, on the exact data that caused the violation. 

\subsection{The Assurance Feedback Loop: Learning from Failure}
Every violation detected at runtime is an invaluable piece of data. It represents a concrete, real-world scenario where the AI model's behavior deviated from its specified contract. FAME institutionalizes the collection of these violation contexts-logging the input data, the AI's erroneous output, and the specific rule that was violated. This data is fed back to engineering teams, creating a powerful, learning-based assurance cycle. This feedback is used to:
\begin{itemize}
    \item \textbf{Improve the AI Model:} The logged failure cases form a highly curated dataset of the model's most critical weaknesses. This data is far more valuable for targeted retraining than millions of randomly collected nominal data points.
    \item \textbf{Refine Formal Specifications:} A pattern of violations might indicate that a specification is too stringent or does not correctly capture the system's operational design domain (ODD). The feedback loop allows for data-informed refinement of the safety contract.
    \item \textbf{Enhance Mitigation Strategies:} Analysis of failure modes can inform the design of more effective mitigation strategies.
\end{itemize}
This process transforms the system from a static entity into one that continuously learns and hardens its safety capabilities throughout its operational life.

\section{Proof of Concept: Securing an Autonomous Perception System}
We validated FAME using a YOLOv4-based pedestrian detection system in the high-fidelity CARLA simulator.

\subsection{Detailed Experimental Setup}
\subsubsection{System Architecture and Model} The simulated vehicle was equipped with a forward-facing camera sensor model. The perception component utilized a pre-trained YOLOv4 DNN, renowned for its robust performance on object detection benchmarks. The baseline system's performance on the COCO dataset was a mean Average Precision (mAP) of 65.7\%. The system ran on a simulated compute stack mimicking a typical automotive-grade ECU.

\subsubsection{Safety Specification and Monitor Synthesis} We established a key safety property derived from functional safety requirements: "Any object classified as a pedestrian within a 30-meter forward cone must be consistently detected with a confidence score above 0.8 for at least 90\% of frames while it remains in the cone." This property was formalized into an STL specification. Using a library based on established automata-theoretic principles, we generated a C++ runtime monitor. The monitor's performance was measured: it consumed less than 0.1\% of the CPU time of the YOLOv4 inference process and had a memory footprint of under \SI{1}{\mega\byte}. 
This demonstrates its suitability for real-time deployment; although overhead scales with specification complexity, it remains negligible for the vast majority of practical safety rules.

\subsubsection{Scenario Generation} We subjected the system to 200 scenarios, split into two categories:
\begin{itemize}
    \item \textbf{Nominal Scenarios (100):} Procedurally generated common driving situations with clear visibility, varied pedestrian paths, and normal traffic, designed to confirm baseline performance.
    \item \textbf{Challenging Scenarios (100):} Specifically crafted to probe known DNN weaknesses. These were generated by systematically varying environmental parameters: rain intensity (from 0 to 100), sun angle to induce severe glare, and fog density. We also programmatically introduced partial occlusions of pedestrians using street furniture.
\end{itemize}
The baseline system's reliability was measured by logging every instance where its output failed to satisfy the STL property. We then deployed the same system with the FAME monitor actively observing the perception output.

\subsection{Expanded Results and In-Depth Analysis}
In the 100 nominal scenarios, the DNN performed with 99\% reliability, and the FAME monitor correctly confirmed this behavior, generating zero false alarms. This confirmed that the monitor does not impede nominal operations.

The results from the 100 challenging scenarios, however, were starkly different. The DNN's performance degraded significantly, failing to meet the safety requirement in \textbf{31 of the 100 runs}. These were all silent failures; the system provided no internal indication of its degraded perception.

\begin{figure}[!t]
\centering
\includegraphics[width=\columnwidth]{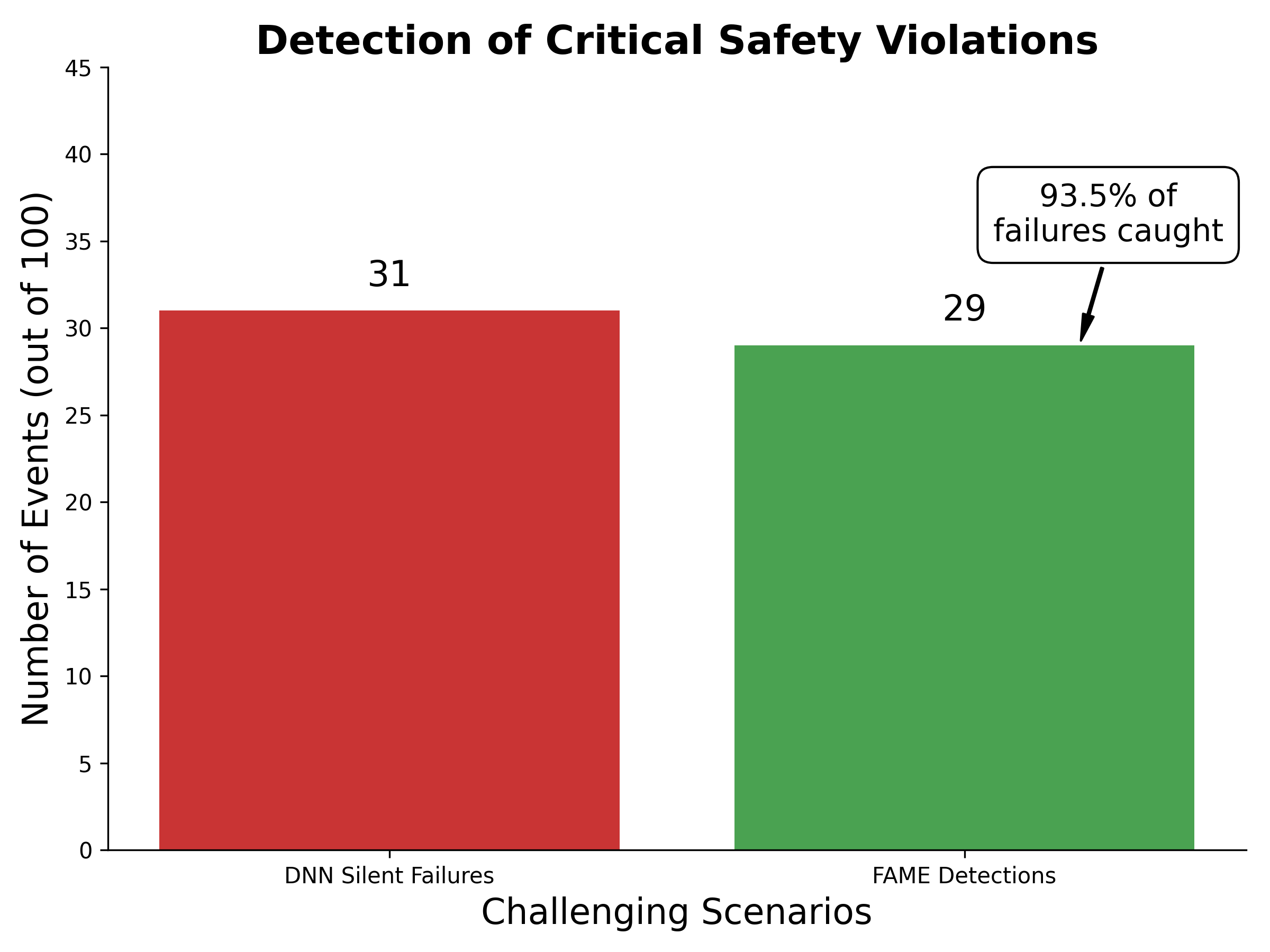}
\caption{Comparison of DNN silent failures versus violations detected by the FAME monitor in 100 challenging scenarios. FAME provided a crucial safety warning in over 93\% of the failure cases.}
\label{fig_results}
\end{figure}

As shown in Fig. \ref{fig_results}, the FAME monitor successfully detected \textbf{29 of the 31 silent failures (a 93.5\% detection rate)}. To understand the significance of this, we performed a qualitative analysis of the detected failures.

\subsubsection{Case Study 1: Failure due to Partial Occlusion} In one scenario, a pedestrian walking on a sidewalk was briefly occluded up to 40\% by a mailbox. The DNN's confidence score, which had been stable at ~0.95, dropped to 0.6 for three consecutive frames before recovering. While the pedestrian was never fully "lost," this flicker violated the consistency requirement of the STL specification. The FAME monitor instantly flagged this transient instability, which could be critical if the vehicle were making a turn decision at that exact moment. The baseline system was completely unaware of this temporary degradation in quality of service.

\subsubsection{Case Study 2: Failure due to Sensor Glare} In another scenario, the simulated vehicle turned a corner directly into a low sun. The camera sensor was saturated with glare for approximately 1.5 seconds. A pedestrian who was clearly visible before and after the turn was completely missed by the DNN for 12 consecutive frames during the glare event. The FAME monitor, which tracks objects over time, detected that a previously tracked pedestrian, who was at a close distance, had vanished from the perception output without exiting the region of interest. It immediately triggered a violation, providing the critical information needed to engage a fail-safe maneuver.

\subsubsection{Analysis of Missed Detections (False Negatives)} The two failures missed by the monitor were as instructive as those it caught. In both cases, the DNN failed to detect the pedestrian but misclassified it as a different object (e.g., a "statue") with high confidence. Our initial STL specification was focused on detection confidence and consistency, and did not include a rule against such semantic misclassifications. This represents a specification gap, not a monitor failure. The data logged from these events would, via the assurance feedback loop, lead directly to the creation of a new, more robust specification: $G ((\text{is\_ped}_{\text{truth}}) \implies \neg (\text{is\_statue}_{\text{AI}}))$. This demonstrates the evolutionary nature of the FAME framework.

\subsubsection{Analysis of False Positives (FP)}
\label{subsec:fp_latency}
A critical metric for any safety monitor is its rate of false positives (FP)—incorrectly flagging a violation during nominal, safe operation. In the context of autonomous systems, a high FP rate leads to "nuisance activations," where the vehicle might perform an unnecessary and potentially hazardous mitigation action, such as sudden braking on a clear highway. This not only creates new risks but also erodes trust and usability.

In the 100 nominal scenarios, which were designed to reflect normal driving conditions, the FAME monitor generated zero false positives. This 100\% specificity is a crucial result, demonstrating that the monitor can provide its safety oversight without interfering with the system's intended function. This outcome is not incidental but a direct result of the FAME methodology. The design-time specification engineering loop—where parameters like confidence thresholds and consistency windows are tuned against both nominal and edge-case data—serves to calibrate the monitor to tolerate acceptable, minor fluctuations in the AI's output while remaining highly sensitive to genuine, safety-critical degradations.

While the observed FP rate was 0 for our 100 trials, to rigorously account for sampling limitations, we calculated the exact two-sided 95\% Clopper-Pearson confidence interval for this result. For n=100 observations with zero events, the confidence interval for the true FP rate is [0,0.036]. This means we can be 95\% confident that the true rate of false positives is no higher than 3.6\%, providing strong statistical evidence of the monitor's reliability in nominal conditions. This result, taken alongside the 93.5\% detection rate in challenging scenarios, shows that a well-parameterized monitor can effectively navigate the fundamental trade-off between detection sensitivity and operational trustworthiness.

\section{Discussion: From Research to Certified Reliability}
The FAME framework is not merely a theoretical construct; it is a practical methodology that aligns with the stringent demands of industrial safety standards. Its primary value lies in its ability to manage and mitigate the residual risk associated with opaque AI components.

\subsection{A Concrete Pathway to Certification: ISO 26262 and ISO/PAS 8800}
For the automotive industry, ISO 26262 \cite{iso26262_2018} is the benchmark for functional safety. A major hurdle for AI-based systems is demonstrating compliance, particularly for functions with high Automotive Safety Integrity Levels (ASILs), such as ASIL D for autonomous driving perception. The standard requires that risks be identified and controlled by independent safety mechanisms with sufficient diagnostic coverage. The more recent ISO/PAS 8800 standard builds on this by specifically addressing the safety of AI in road vehicles, emphasizing the need for processes to manage the risks associated with learning-enabled systems, including their unpredictable behavior with OOD inputs. This section discusses how the FAME framework provides a concrete pathway to satisfying the requirements of both standards.

\subsubsection{ISO 26262: Functional Safety}
Under ISO 26262, the FAME monitor acts as a verifiable \textbf{safety mechanism}. It can be used to detect failures in the AI component (the primary function) and trigger a transition to a safe state. This aligns perfectly with the concept of \textbf{ASIL decomposition}. A complex, hard-to-verify AI component (responsible for an ASIL D function) can be paired with a simpler, formally synthesized FAME monitor (verifiable to a lower level, e.g., ASIL B) and a verified safe state mechanism. This decomposition enables engineers to establish a robust safety case, arguing that even if the complex AI component fails silently, the FAME safety mechanism provides the necessary diagnostic coverage (detecting the failure) and risk mitigation (triggering the safe state) to meet the overall safety goals. This provides a clear, defensible, and auditable path for certifying AI systems for use in the most safety-critical applications. 

\subsubsection{ISO/PAS 8800: Safety and Artificial Intelligence}
ISO/PAS 8800 directly confronts the unique challenges of AI, particularly the unpredictability of learning-enabled components when faced with novel or OOD inputs. FAME provides a powerful solution by establishing a concrete mechanism to \textbf{define and enforce explicit behavioral boundaries} on the AI component. The formal specifications synthesized into a FAME monitor act as a deterministic safety net, ensuring the AI's output remains within a pre-defined safe operating envelope, regardless of its internal state or input novelty. Furthermore, the framework's assurance feedback loop, where runtime violations are logged and used to refine the AI model or its safety specifications, directly supports the standard's call for \textbf{continuous monitoring and improvement} throughout the vehicle's lifecycle. This transforms the abstract requirement of managing AI-specific risks into a practical, verifiable engineering process. Practically, FAME contributes to ISO/PAS 8800 clauses on operational monitoring, risk control, and human oversight by providing traceable links from safety goals to formal specifications to runtime verdicts and audit logs.

\section{Cross-Domain Design Patterns}
FAME's methodology formalizes contracts over observable signals, synthesizes bounded-future monitors with RTAMT, and enforces mitigations, \emph{conceptually} extends beyond automotive. We present patterns in Table~\ref{tab:domain_apps}; empirical validation remains an item of future work.

\begin{table*}[!ht]
\caption{Cross-Domain Design Patterns of the FAME Framework}
\label{tab:domain_apps}
\centering
\begin{tabular}{@{}p{2.7cm}p{5.2cm}p{8.1cm}@{}}
\toprule
\textbf{Domain} & \textbf{Example Silent Failure} & \textbf{Example FAME Monitor Specification} \\ \midrule
Medical AI & Chest X-ray CAD model confidently misses a small yet suspicious nodule. & If (nodule\_features\_present) $\wedge$ (nodule\_size $>$ \SI{5}{\milli\meter}), then within \SI{1}{\second} (malignancy\_conf $>$ 0.95) \textbf{OR} route to a safety-certified rule-based second-opinion model; escalate only if both disagree persistently. \\ \addlinespace
Industrial Robotics & Vision mislocalizes a part due to reflections, causing a near-collision. & End-effector trajectory shall never intersect defined safety zones (geometric guard over planned path). \\ \addlinespace
Aerospace UAS\footnote{UAS: Unmanned Aircraft System.} & Terrain/altitude misinterpretation violates airspace limits. & GPS/baro-derived state must remain within corridor/geofence at all times. \\ \addlinespace
Critical Infrastructures & AI load balancer over-commands a feeder, risking cascading trips. & Over any window $\Delta t$, energy out of a node must not exceed measured/estimated energy in plus storage delta. \\ \addlinespace
Financial Systems & Algorithmic trader rapidly exceeds risk budget. & 5-minute rolling exposure of AI-initiated orders shall not exceed the limit $X$; violation triggers throttling. \\ \addlinespace
Cybersecurity & DNN scanner labels code "safe," missing an SQL injection \cite{app14188365}. & If AI labels "safe," it must also pass a lightweight static pattern check for critical injection syntax; otherwise, block. \\ \bottomrule
\end{tabular}
\end{table*}

\section{Empirical Scope and Threats to Validity}
\label{sec:scope}
Our empirical claims are scoped to a single domain: a simulated autonomous driving perception stack with a single STL property family and a single model type (YOLOv4). Cross-domain items in Table~\ref{tab:domain_apps} are patterns meant to guide specification engineering, not proof of generalizability. External validity is bounded by: (i) simulator realism and scenario coverage; (ii) choice of thresholds/windows; and (iii) reliance on observable signals (hidden internal failures that do not manifest at the interfaces may evade detection). These limits motivate the design-time stressing loop and the lifecycle feedback process. While these cross-domain patterns are conceptual, they provide a concrete starting point for reliability engineers to adapt the FAME methodology to their specific contexts, bridging the gap between theory and diverse industrial applications.




\section{Limitations and Future Research}
No framework is a panacea. FAME's primary limitation is that it is only as good as its specifications. It cannot protect against "unknown unknowns" if those scenarios are not covered by any safety rule. Future work will expand property mining, multi-sensor guards, and compositional assurance across perception–prediction–planning. Empirical replication in robotics and medical imaging is planned to complement the current case study.

\section{Conclusion and Future Vision}
The adoption of AI in critical systems necessitates the engineering of a new class of silent failures. Probabilistic models may fail; our task as reliability engineers is not to prevent every failure, but to ensure that when they do fail, they do so safely. The FAME framework provides a robust, practical, and extensible solution by combining the foresight of formal synthesis with the vigilance of runtime enforcement. Our proof-of-concept demonstrates that this approach can effectively capture dangerous AI behaviors that elude traditional verification and validation, offering a crucial layer of defense.

The journey, however, is not over. We see several exciting frontiers for this research:
\begin{itemize}
    \item \textbf{Generative Assurance:} Integrating monitor synthesis directly into the MLOps pipeline, potentially using Large Language Models to help engineers draft initial STL specifications from natural language requirements, which are then formally refined and synthesized.
    \item \textbf{Self-Adapting Monitors:} Creating monitors that can learn from benign violations to reduce false positives or automatically suggest refinements to their own specifications based on operational data, creating a truly self-improving assurance system.
    \item \textbf{Compositional Assurance:} Developing theories and tools for composing monitors across multiple AI components (e.g., perception, prediction, and planning) to provide end-to-end, system-level safety guarantees with verifiable properties.
\end{itemize}
By pursuing these avenues, we can move beyond the current paradigm of best-effort, probabilistically-correct AI. FAME provides a foundational step towards a future where AI systems are not only powerful and intelligent but are also provably and demonstrably reliable. It offers a path to deploy advanced AI not with blind faith, but with formally assured confidence.

\section{ACKNOWLEDGMENTS}
The authors would like to thank Professor Kuo-Hui Yeh at National Yang Ming Chiao Tung University, the editor, and the anonymous reviewers for their invaluable feedback and suggestions, which greatly improved the quality of this paper. This work was supported in part by the National Science and Technology Council, Taiwan R.O.C., under NSTC 114-2221-E-002-217, NSTC 114-2622-E-A49-022, NSTC 114-2221-E-A49-210, NSTC 114-2634-F-011-002-MBK, and NSTC 114-2923-E-194-001-MY3. Additionally, this work received partial funding from the National Taiwan University under Grant G0647 and Grant 114L895501, within the NTU Core Consortium Project by National Taiwan University and the framework of the Higher Education Sprout Project by the Ministry of Education, Taiwan. Additional partial financial support was provided by the Department of Industrial Technology, Ministry of Economic Affairs, under the “2025 ITRI Advanced Research Program” (Grant No.: 114-EC-17-A-21-0337). Further partial support was also granted by the Hon Hai Research Institute in Taipei, Taiwan (Project No.: 114UA90042), and the Industry-Academia Innovation School, NYCU, Taiwan (Project No.: 113UC2N006). The authors would like to express their gratitude for the financial support mentioned above.  
Guan-Yan Yang is grateful to the National Science and Technology Council (NSTC) in Taiwan for the graduate research fellowship (NSTC-GRF) and to Professor Hung-Yi Lee for co-hosting his Ph.D research project. 
He also thanks the research scholarship from the Norman and Lina Chang Foundation in California, USA.

\def\refname{REFERENCES}
\bibliographystyle{IEEEtran}
\bibliography{ref}

\begin{thebibliography}{10}
\providecommand{\url}[1]{#1}
\csname url@samestyle\endcsname
\providecommand{\newblock}{\relax}
\providecommand{\bibinfo}[2]{#2}
\providecommand{\BIBentrySTDinterwordspacing}{\spaceskip=0pt\relax}
\providecommand{\BIBentryALTinterwordstretchfactor}{4}
\providecommand{\BIBentryALTinterwordspacing}{\spaceskip=\fontdimen2\font plus
\BIBentryALTinterwordstretchfactor\fontdimen3\font minus \fontdimen4\font\relax}
\providecommand{\BIBforeignlanguage}[2]{{%
\expandafter\ifx\csname l@#1\endcsname\relax
\typeout{** WARNING: IEEEtran.bst: No hyphenation pattern has been}%
\typeout{** loaded for the language `#1'. Using the pattern for}%
\typeout{** the default language instead.}%
\else
\language=\csname l@#1\endcsname
\fi
#2}}
\providecommand{\BIBdecl}{\relax}
\BIBdecl

\bibitem{hendrycks2018baselinedetectingmisclassifiedoutofdistribution}
\BIBentryALTinterwordspacing
D.~Hendrycks and K.~Gimpel, ``A baseline for detecting misclassified and out-of-distribution examples in neural networks,'' 2018. [Online]. Available: \url{https://arxiv.org/abs/1610.02136}
\BIBentrySTDinterwordspacing

\bibitem{silentfailure2023}
R.~Karval and K.~N. Singh, ``Catching silent failures: A machine learning model monitoring and explainability survey,'' in \emph{2023 OITS International Conference on Information Technology (OCIT)}, 2023, pp. 526--532.

\bibitem{zhu2023verification}
Q.~Zhu, W.~Li, C.~Huang, X.~Chen, W.~Zhou, Y.~Wang, J.~Li, and F.~Fu, ``Verification and design of robust and safe neural network-enabled autonomous systems,'' in \emph{2023 59th Annual Allerton Conference on Communication, Control, and Computing (Allerton)}.\hskip 1em plus 0.5em minus 0.4em\relax IEEE, 2023, pp. 1--8.

\bibitem{cheng2023runtime}
C.-H. Cheng, M.~Luttenberger, and R.~Yan, ``Runtime monitoring dnn-based perception: (via the lens of formal methods),'' in \emph{International Conference on Runtime Verification}.\hskip 1em plus 0.5em minus 0.4em\relax Springer, 2023, pp. 428--446.

\bibitem{yang2025resilience}
G.-Y. Yang, J.-N. Chen, F.~Wang, and K.-H. Yeh, ``Enhancing resilience for {IoE}: A perspective of networking-level safeguard,'' \emph{IEEE Network}, pp. 1--11, 2025.

\bibitem{bartocci2018survey}
\BIBentryALTinterwordspacing
E.~Bartocci, J.~Deshmukh, A.~Donz{\'e}, G.~Fainekos, O.~Maler, D.~Ni{\v{c}}kovi{\'{c}}, and S.~Sankaranarayanan, \emph{Specification-Based Monitoring of Cyber-Physical Systems: A Survey on Theory, Tools and Applications}.\hskip 1em plus 0.5em minus 0.4em\relax Cham: Springer International Publishing, 2018, pp. 135--175. [Online]. Available: \url{https://doi.org/10.1007/978-3-319-75632-5_5}
\BIBentrySTDinterwordspacing

\bibitem{cofer2020run}
D.~Cofer, I.~Amundson, R.~Sattigeri, A.~Passi, C.~Boggs, E.~Smith, L.~Gilham, T.~Byun, and S.~Rayadurgam, ``Run-time assurance for learning-enabled systems,'' in \emph{NASA Formal Methods Symposium}.\hskip 1em plus 0.5em minus 0.4em\relax Springer, 2020, pp. 361--368.

\bibitem{sun2025correct}
M.~Sun, G.~Bakirtzis, H.~Jafarzadeh, and C.~Fleming, ``Correct-by-construction requirement decomposition,'' \emph{Software and Systems Modeling}, pp. 1--16, 2025.

\bibitem{wang2004formal}
F.~Wang, ``Formal verification of timed systems: A survey and perspective,'' \emph{Proceedings of the IEEE}, vol.~92, no.~8, pp. 1283--1305, 2004.

\bibitem{iso26262_2018}
\BIBentryALTinterwordspacing
{International Organization for Standardization}, \emph{Road vehicles -- Functional safety}, International Organization for Standardization Std. ISO 26\,262, 2018. [Online]. Available: \url{https://www.iso.org/standard/68385.html}
\BIBentrySTDinterwordspacing

\bibitem{iso8800_2024}
\BIBentryALTinterwordspacing
------, \emph{Road vehicles -- Safety and artificial intelligence}, International Organization for Standardization Std. ISO/PAS 8800:2024, 2024. [Online]. Available: \url{https://www.iso.org/standard/83303.html}
\BIBentrySTDinterwordspacing

\bibitem{wang2013temporal}
F.~Wang, J.-H. Wu, C.-H. Huang, C.-C. Chang, and C.-C. Li, ``Temporal specification mining for anomaly analysis,'' in \emph{Asian Symposium on Programming Languages and Systems}.\hskip 1em plus 0.5em minus 0.4em\relax Springer, 2013, pp. 273--289.

\bibitem{kuznietsov2024explainable}
A.~Kuznietsov, B.~Gyevnar, C.~Wang, S.~Peters, and S.~V. Albrecht, ``Explainable ai for safe and trustworthy autonomous driving: A systematic review,'' \emph{IEEE Transactions on Intelligent Transportation Systems}, 2024.

\bibitem{atakishiyev2025safety}
S.~Atakishiyev, M.~Salameh, and R.~Goebel, ``Safety implications of explainable artificial intelligence in end-to-end autonomous driving,'' \emph{IEEE Transactions on Intelligent Transportation Systems}, 2025.

\bibitem{app14188365}
\BIBentryALTinterwordspacing
G.-Y. Yang, F.~Wang, Y.-Z. Gu, Y.-W. Teng, K.-H. Yeh, P.-H. Ho, and W.-L. Wen, ``{TPSQLi}: Test prioritization for sql injection vulnerability detection in web applications,'' \emph{Applied Sciences}, vol.~14, no.~18, 2024. [Online]. Available: \url{https://www.mdpi.com/2076-3417/14/18/8365}
\BIBentrySTDinterwordspacing

\end{thebibliography}

\vspace*{-8pt}


\begin{IEEEbiography}{Guan-Yan Yang}{\,}received a Bachelor's degree from the Department of Information Management at National Dong Hwa University, Hualien, Taiwan, in 2022. 
    He is currently pursuing a Ph.D. in the Department of Electrical Engineering at National Taiwan University, Taipei, Taiwan.
    In 2023, he worked as a Software Engineer Intern at the R\&D-DTP of TSMC. Since 2024, he has been a researcher at the Taiwan Academic Cybersecurity Center, NTUST. In 2024, he was awarded a scholarship from the Norman and Lina Chang Foundation in the USA.
    His research interests include security, safety, deep learning, generative AI, the Internet of Things, formal verification, and software testing. He is a member of the IEEE Computer Society, the IEEE Reliability Society, the IEEE Communications Society, the IEEE Consumer Technology Society, SEAT, and the ACM SIGMICRO. Contact him at f11921091@ntu.edu.tw. 
\end{IEEEbiography}

\begin{IEEEbiography}{Farn Wang}{\,}is a Full Professor at the Department of Electrical Engineering, National Taiwan University.
    He received the B.S. degree in Electrical Engineering from National Taiwan University in 1982 and the M.S. degree in Computer Engineering from National Chiao-Tung University in 1984. 
    He completed his Ph.D. in Computer Science at the University of Texas at Austin in 1993. 
    He is a founding member and chairman of the Steering Committee of the International Symposium on Automated Technology for Verification and Analysis (ATVA) from 2003 to 2022, and has served on the ATVA advisory committee since 2022.  
    He was also an Associate Editor of FMSD (International Journal on Formal Methods in System Design), Springer-Verlag.   
    His research interests include formal verification, model-checking, software testing, security, verification automation, AI, and language models.
    He has been named a World's Top 2\% Scientists in a career-long list by Stanford University since 2020.
\end{IEEEbiography}

\end{document}